\documentclass[prc,floatfix,onecolumn]{revtex4}
\usepackage{color}
\usepackage{amsfonts}
\usepackage{amsbsy}
\usepackage{mathrsfs}
\usepackage{graphicx}
\usepackage{subfigure}
\newcommand{\eeq}{\end{equation}}
\newcommand{\bea}{\begin{eqnarray}}
\newcommand{\eea}{\end{eqnarray}}

\def\lsim{\mathrel{\rlap{
\lower4pt\hbox{\hskip-3pt$\sim$}}
    \raise1pt\hbox{$<$}}}     %less than approx. symbol
\def\gsim{\mathrel{\rlap{
\lower4pt\hbox{\hskip-3pt$\sim$}}
    \raise1pt\hbox{$>$}}}     %greater than or approx. symbol

\begin{document}

%\DOIsuffix{theDOIsuffix}

%\Volume{XX}
%\Month{XX}
%\Year{XXXX}

%\pagespan{1}{}

%\Receiveddate{XXXX}
%\Reviseddate{XXXX}
%\Accepteddate{XXXX}
%\Dateposted{XXXX}

\title{Color path integral Monte Carlo simulations \\ %of strongly coupled 
of quark-gluon plasma%\\ Thermodynamic properties.
%  \footnote{Dedicated to   A.B. Migdal on the occasion of his 100th birthday.}
  }%
%}
\author{V.S.~Filinov}
\thanks{Corresponding author\quad E-mail:~\textsf{vladimir\_filinov@mail.ru}}
\affiliation{Joint Institute for High Temperatures, Russian Academy of
Sciences, Moscow, Russia}
\author{Yu.B. Ivanov}%{e-mail: Y.Ivanov@gsi.de}
%\affiliation{GSI Helmholtzzentrum
% f\"ur Schwerionenforschung, % GmbH,
%Planckstr. 1, D-64291
%Darmstadt, Germany}
\affiliation{Kurchatov Institute,
Kurchatov sq. 1,
Moscow, Russia}
%\author[]{Y.B.~Ivanov\inst{3,4}}
%\author[]{V.V.~Skokov\inst{3,5}}
\author{M. Bonitz}
\affiliation{Institute for Theoretical Physics and Astrophysics, Christian Albrechts University, % Kiel,
% Leibnizstrasse 15, D-24098
Kiel, Germany}
\author{V.E. Fortov}
\affiliation{Joint Institute for High Temperatures, Russian Academy of
Sciences, Moscow, Russia}
\author{P.R. Levashov}
\affiliation{Joint Institute for High Temperatures, Russian Academy of
Sciences, Moscow, Russia}
%
%---------------------------


\begin{abstract}
Thermodynamic properties of a 
strongly coupled quark-gluon plasma (QGP) of constituent quasiparticles
is studied by a color path-integral Monte-Carlo simulations (CPIMC). 
For our simulations we have presented QGP partition function in the form of color path integral with new relativistic measure instead of Gaussian one used in Feynman and Wiener path integrals. For integration over color variable we have also  developed procedure of sampling color variables according to the group SU(3) Haar measure. It is shown that this method is able to reproduce the available quantum lattice chromodynamics (QCD)  data.  
\end{abstract}

%\pacs{12.38Mh, 31.15.Qg, 51.20.+d, 52.27Gr}
%\keywords{strongly correlated plasma, quark gluon plasma}

\maketitle

\section{Introduction}\label{s:intro}

%Contributions of A.B. Migdal to physics of the last century  are difficult to overestimate.
%These include the Landau-Migdal-Pomeranchuk effect, theory of finite Fermi systems,
%pionic degrees of freedom in nuclei, physics of strong fields, {\it etc}.  This results entered many
%textbooks and found numerous applications. A.B. Migdal was one of the first who started to think
%about possible existence of an anomalous nuclear matter \cite{Migdal}. Nowadays this concept is, in particular,
%associated with the phase transition into the QGP. While the possibility of pion
%condensation is still vividly discussed in astrophysical applications, see, e.g., \cite{Voskresensky}.

Investigation of properties of the QGP is
one of the main challenges of strong-interaction physics, both
theoretically
%\cite{ZW}
and experimentally.
%\cite{GT}.
Many features of this matter were
experimentally discovered at the Relativistic Heavy Ion Collider
(RHIC) at Brookhaven. The most striking result, obtained from analysis
of these experimental data \cite{shuryak08}, is that the deconfined
quark-gluon matter behaves as an almost perfect fluid rather than a perfect gas,
as it could be expected from the asymptotic freedom.
There are  various theretical approaches to studying QGP.
Each approach has its advantages and  disadvantages.
The most fundamental way to compute
properties of the strongly interacting matter is provided by the lattice QCD \cite{%Lattice09,
Fodor09,Csikor:2004ik}.
Interpretation of these very complicated Monte Carlo computations
requires application of various QCD motivated, albeit schematic, models simulating various aspects of the full theory.
%and allowing for a deeper physical understanding.
Moreover, such models are needed in cases when the lattice QCD fails, e.g. at large
baryon chemical potentials and out of equilibrium.
While some progress has been achieved in the recent years,
we are still far away from having a satisfactory understanding of the QGP dynamics.
%It is therefore mandatory to devise reliable and manageable theoretical tools for a quantitative description of non-Abelian QGP both in- and out--of equilibrium.

A semi-classical approximation, based on a point like quasiparticle picture, has been introduced in \cite{Levai,LM105,LM109,LM110,LM111}.
It is expected that the main features of non-Abelian
plasmas can be understood in simple semi-classical terms without
the difficulties inherent to a full quantum field theoretical analysis.
Independently the same ideas were implemented in terms of molecular dynamics (MD) \cite{Bleicher99}.
Recently this MD approach was further developed in a series of works
\cite{shuryak1,Zahed}. The MD allowed one to treat soft processes in the QGP which
are not accessible by perturbative means.

%This article reviews in detail the conceptual framework for
%semi-classical approximation for strongly coupled non-Abelian plasma within Feynman and Wigner formulation of quantum %mechanics.

A strongly correlated behavior of the QGP is expected to show up in long-ranged spatial correlations of quarks and
gluons which, in fact, may give rise to liquid-like and, possibly, solid-like structures.
This expectation is based on a very similar
behavior observed in electrodynamic plasmas  \cite{thoma04,afilinov_jpa03}. This similarity was exploited
to formulate a classical nonrelativistic model of a color Coulomb interacting QGP \cite{shuryak1} which was  numerically analyzed by classical MD simulations.
Quantum effects were either neglected or included
phenomenologically via a short-range repulsive correction to the pair potential. Such a rough model may become
a critical issue at high densities,
where quantum effects strongly affects properties of the QGP.
%Similar models have been used in
%electrodynamic plasmas and showed poor behavior in the region of strong wave function overlap, in particular at the Mott density.
%For temperatures and densities of the QGP considered
%in  \cite{shuryak1} these effects are very important as the quasiparticle
%thermal wave length is of order the average interparticle distance.
%This difficulty can be eliminated by deriving effective quantum potentials, as shown by some of the present authors before
%\cite{afilinov_jpa03,afilinov_pre04,ebeling06}. Following an idea of Kelbg \cite{kelbg}, quantum corrections to the pair potential
%can be rigorously derived in perturbation theory with respect to the coupling parameter \cite{dusling09}.
%Beside the
%nonideality and quantum effects our approach  takes into account the effects of the Fermi (Bose) statistics of quarks (gluons)
%by a proper antisymmetrization (symmetrization) of the $N-$body density matrix.
To account for the quantum effects we follow an idea of Kelbg \cite{kelbg} rigorously allowing for quantum corrections to the pair potential %are derived 
\footnote{
The idea to use a Kelbg-type effective potential also for quark matter
was independently proposed  by K.~Dusling
and C.~Young \cite{dusling09}. However, their potentials are limited to weakly nonideal systems.
}. 
%The main goal is to test this approach for ability to reproduce the equation of state known from lattice data \cite{Lattice09,Fodor09}.
%To this end we use the simplest model of a QGP consisting of quarks, antiquarks and gluons interacting via a color Coulomb potential due to
%Gelman et al. \cite{shuryak1}.
% with several approximations for the temperature dependence of the quasiparticle masses.
%We report surprisingly
%good agreement with the lattice data for one of the parameter sets, which gives us confidence that the model correctly captures main properties
%of the nonideal QGP. 
To extend the method of quantum effective potentials to a stronger coupling, we 
%
%an ``improved Kelbg potential''
%was derived, which contains a single free parameter,
%being fitted to the exact solution of the quantum-mechanical two-body problem.
%Thus, the method of the improved Kelbg potential is able to describe
%thermodynamic properties up to moderate couplings \cite{afilinov_pre04}. 
%
%However, this approach may fail, if bound states of more than two
%particles are formed in the system. This results in
%a break-down of the pair approximation for the density matrix, as demonstrated in  \cite{afilinov_pre04}. A superior %approach, which does not have this limitation, consists in 
%
use the original Kelbg potential in the path integral approach, which effectively map the problem onto a high-temperature weakly coupled and weakly degenerate one. 
This allows one to extend the analysis to strong couplings and is, therefore, a relevant choice for the present purpose.
%
%This method has been successfully applied to strongly coupled electrodynamic plasmas
%\cite{filinov_ppcf01,bonitz_jpa03}. %,bonitz_pop08}.
%Examples are partially-ionized dense hydrogen plasmas, where liquid-like and
%crystalline behavior was observed \cite{filinov_jetpl00,bonitz_prl05}.
%Moreover, also partial ionization effects and pressure ionization
%could be studied from first principles \cite{filinov_jpa03}.
%The same methods have been also applied to electron-hole plasmas in
%semiconductors \cite{bonitz_jpa06,filinov_pre07}, including excitonic bound states,
%which have many similarities to the QGP due to
%smaller mass differences as compared to electron-ion plasmas.

In this paper we 
%extend the analysis to dynamic properties by performing
%first exploratory quantum molecular dynamics (QMD) simulations of the QGP.
%This MD method
extend previous classical nonrelativistic simulations \cite{shuryak1}
based on a color Coulomb interaction to the quantum regime.
For quantum Monte Carlo simulations of the thermodynamic properties of QGP we have to rewrite the partition function of this system in the form of color path integrals with a {\em new relativistic measure} instead of Gaussian one used in Feynman and Wiener path integrals. For integration partition function over color variables we have   developed procedure of sampling color quasiparticle variables according to {\em the group SU(3) Haar measure} with the quadratic and cubic Casimir conditions. The developed approach self-consistently takes  into account the Fermi (Bose) statistics of quarks (gluons).
The main goal of this article is to test the developed approach for ability to reproduce known lattice data \cite{%Lattice09,
Fodor09} and to predict other properties of the QGP, which are still unavailable for the
lattice calculations. 
%To this end we have developed quantum generalization of a simple model %\cite{shuryak1} of the QGP consisting of quarks, antiquarks and gluons interacting via a color Coulomb potential.
First results of applications of the path integral approach to study of thermodynamic 
properties of the nonideal QGP have already been briefly reported in 
\cite{Filinov:2009pimc,Filinov:2010pimc}. In this paper we have shown that CPIMC is
able to reproduce the QCD lattice equation of state and related thermodynamic functions 
% as , internal energy, entropy and trace anomaly 
and also yields valuable insight into the internal structure of the QGP. 

Theoretical treatment and hydrodynamic simulations of the experimentally observed 
expansion of the fireball consisting  of quarks and gluons at relativistic heavy-ion collisions suggests 
knowledge not only thermodynamic but also transport properties of the QGP.  
Unfortunately the CPIMC method itself is not able to yield transport properties. 
To simulate quantum QGP transport and thermodynamic properties 
within unified approach we are going to combine path integral and Wigner (in phase space) formulations 
of quantum mechanics. The canonically averaged quantum operator time correlation functions and related kinetic 
 coefficients will be calculated according to the Kubo formulas. In this approach CPIMC is used 
 not only for calculation thermodynamic functions but also to generate initial conditions 
 (equilibrium spatial, momentum, spin, flavor and color quasiparticle configurations) for generation 
of the color trajectories being the solutions of related differential dynamic equations. 
% described the time evolution for spatial, momentum and color variables. 
Correlation functions and kinetic coefficients are calculated as averages  
of Weyl's symbols of dynamic operators along these trajectories. 
%May be to correct this calculation something like color friction should be included 
%in dynamic equations.  
The basic ideas of this approach have been published in \cite{ColWig11}. 
Using this approach we are going to calculate diffusion coefficient and  viscosity of the strongly coupled QGP. 
 Work on these problems is in progress. In this dynamic approach energy of the 
quasiparticle system is conserved. 
 In contrast known attempts to use quantum discreet color variable in classical simulations combining dynamic differential equations for quasiparticle coordinates and momenta  and discrete color variables results in 
violation of energy conservation low due to unphysical increase of kinetic energy \cite{Levai}. 

%----------------------------
\section{Thermodynamics of QGP} \label{Thermodynamics}
\subsection{Basics of the model}\label{semi:model}

Our model is based on a resummation technique and lattice simulations for %allowing for consideration of the QGP
%as system of
dressed quarks, antiquarks and gluons interacting via the color Coulomb potential.
The assumptions of the model are similar to those of  \cite{shuryak1}:
% and are summarized as follows :
%
\begin{description}
 \item[I.]% All color quasiparticles are %enough heavy to be
 %more or less localized. This means that their
 Quasiparticles masses ($m$) are of order or higher
 than the mean kinetic energy  per particle. This assumption is based on the analysis of lattice
 data \cite{Lattice02,LiaoShuryak}. For instance, at zero net-baryon density it
 amounts to $m \geq T$, where $T$ is a temperature.
 %\item[II.] To take into account relativistic effects the kinetic part of the full
 %quasiparticles energy is described by the relativistic expression.
 \item[II.] %Since the particles are heavy,
 In view of the first assumption,
 interparticle interaction is dominated
 by a color-electric Coulomb potential, see Eq. (\ref{Coulomb}).
 Magnetic effects are neglected as subleading ones. % in the nonrelavistic limit.
 \item[III.] Relying on the fact that the color representations are large, the color operators %$t^a$
 are substituted by their average values, i.e. by Wong's classical color vectors (8D in SU(3))
 with the quadratic and cubic Casimir conditions \cite{Wong}.
 %(The time evolution of color vectors will be described  in next paper by Wong's dynamics).
 \item[IV.] We consider the 3-flavor quark model. Since the masses of the 'up', 'down'
 and 'strange' quarks  extracted from lattice data are still indistinguishable, we assume
 these masses to be equal.
 %So for quark and antiquark quasiparticles we assume
 %$m^{\tilde{f}}_q=m^{\tilde{f}'}_{\bar{q}} (\tilde{f},\tilde{f}'=$
 % 'up', 'down' and 'strange') and their fractions ($1/3$) to be equal. We account for the quark
 %color and flavors in the proper consideration of Fermi statistics effects. Let us stress
 %that gluon quasiparticles obey the Bose statistics
 %and their masses are independent from masses of quark quasiparticles.
 As for the gluon quasiparticles, we allow their mass to be different (heavier) from that of quarks.
\end{description}
The quality of these approximations and their limitations were discussed in  \cite{shuryak1}.
Thus, %similarly to  \cite{shuryak1}
this model requires the following quantities as a function of temperature and chemical potential
as an input:
\begin{description}
%\begin{enumerate}
\item[1.] the quasiparticle masses, $m$, and
\item[2.] the coupling constant $g^2$.
%see Eq. (\ref{Coulomb}),
%[Note that, because of the running coupling in the QCD, $g^2$ generally depends on $r$ and $T$].
%\item the particle density.
%F\end{enumerate}
\end{description}
All the input quantities should be deduced from the lattice data or from an appropriate model simulating these data.
%-----------------------
\subsection{Color Path Integrals}\label{s:pimc}

Thus,  we consider a multi-component QGP consisting of $N$ color quasiparticles representing $N_g$ dressed gluon and of various flavors $N_q$ dressed quarks and $\bar{N}_{q}$ antiquarks. The Hamiltonian of this system is
 $\hat{H}=\hat{K}+\hat{U}^C$ with
 the kinetic and color Coulomb interaction parts
\begin{eqnarray}
\label{Coulomb}
{\hat{K}}=\sum_i \sqrt{p^2_i+m^2_i(T,\mu_q)}
\qquad
{\hat{U}^C}=\frac{1}{2}\sum_{i\neq j}
%\frac{g^2(|r_i-r_j|,T,\mu_q)
\frac{g^2(T,\mu_q)
\langle Q_i|Q_j \rangle}{4\pi| r_i- r_j|},
\end{eqnarray}
Here $i=1,\ldots,N$ is the quasiparticle index, $N=N_q+\bar{N}_q+N_g$, 
$i$ and $j$ summations run over quark and gluon quasiparticles, 
3D vectors $r_i$ are quasiparticle spatial coordinates, the $Q_i$ denote the Wong's color variable
(8D-vector in the $SU(3)$ group), $T$ is the
temperature,
%$\beta=1/T$
and $\mu_q$ is the quark chemical potential,
$\langle Q_i|Q_j \rangle$ denote scalar product of color vectors. We use relativistic kinematics, as seen from Eq. (\ref{Coulomb}). Nonrelativistic approximation for potential energy is used to disregard magnetic interaction and
retardation in the Coulomb interaction. In fact, the quasiparticle mass and the coupling constant,
as deduced from the lattice data, are functions of $T$ and, in general, $\mu_q$.

The thermodynamic properties in the grand canonical ensemble with given temperature $T$,
baryon ($\mu_q$) and strange ($\mu_s$) chemical potentials,  and fixed volume $V$ are fully described by the
grand partition function\footnote{Here we explicitly write sum over different quark flavors (u,d,s).
Below we will assume that the sum quark degrees of freedom is understood in the same way.
}
\begin{eqnarray}\label{Gq-def}
Z\left(\mu_q,\mu_s,\beta,V\right)
&=&
%\nonumber\\&&
%\sum_{N_q,\bar{N}_{q},N_g}exp(\beta(\mu_q N_q+\mu_{\bar{q}}\bar{N}_{q}|_{\mu_{\bar{q}}=-\mu_q}+\mu_g N_g)|_{\mu_g=0})
%Z(N_q,N_{ \bar{q}},N_g,V;\beta) =
\sum_{N_u,N_d,N_s,\bar{N}_{u},\bar{N}_{d},\bar{N}_{s},N_g}
\frac{\exp\{\mu_q(N_q-\bar{N}_{q})/T\}\;\exp\{\mu_s(N_s-\bar{N}_{s})/T\}}%
{N_u!\;N_d!\;N_s!\;\bar{N}_{u}!\;\bar{N}_{d}!\;\bar{N}_{s}!\;N_g!}
Z\left(N,V,\beta\right) \text{,} 
\cr
%\nonumber\\
%&\times&
&&Z\left(N,V,\beta\right)=\sum_{\sigma} \int\limits_V
dr\; d \mu Q \;\rho(r,Q, \sigma; N_u,N_d,N_s,\bar{N}_{u},\bar{N}_{d},\bar{N}_{s},N_g; \beta),
\end{eqnarray}
where $N_q=N_u+N_d+N_s$ and $\bar{N}_{q}=\bar{N}_{u}+\bar{N}_{d}+\bar{N}_{s}$ are total numbers of quarks and antiquarks of all  flavours, respectively, 
$N_g$ is the number of gluon quasiparticles, 
$\rho(r,Q, \sigma; N_u,N_d,N_s,\bar{N}_{u},\bar{N}_{d},\bar{N}_{s},N_g; \beta)$ denotes the diagonal matrix
elements of the density matrix operator
${\hat \rho} = \exp (- \beta{\hat H})$ and $\beta=1/T$.
Here $\sigma$, $r$ and $Q$ denote the multy-dimensional vectors related to spin, spatial and color degrees of freedom of all quarks, antiquarks and gluons.
%while $f$ is the multy-dimensional vector of the  flavours of
% $N_u,N_d,N_s$ quarks and $\bar{N}_{u},\bar{N}_{d},\bar{N}_{s}$ antiquarks in the ensemble
%respectively. Components of  flavour vector $f$ can take the following values: $f_i='up', 'down', 'strange'$ for
%$i=1,\ldots, N_q+\bar{N}_q$. 
The $\sigma$ summation and integration $dr d \mu Q$ run over all individual degrees of freedom of the particles, 
$d\mu Q$ means differential of the group $SU(3)$ Haar measure. 
Usual choice of the strange  chemical potential is $\mu_s=0$ (nonstrange matter).
Therefore, below we omit $\mu_s$ from the list of variables.

Since the masses and the coupling constant depend on the temperature and baryon chemical potential,
special care should be taken to preserve thermodynamic consistency of this approach.
In order to preserve the thermodynamic consistency,
thermodynamic functions such as pressure, $P$, entropy, $S$, baryon number, $N_B$, and
internal energy, $E$, should be calculated through respective derivatives of
 the logarithm of the partition function
%%
%\begin{eqnarray}
%\label{p_gen}
%P&=&\partial (T\ln Z) / \partial V, %&\quad&
%\\
%\label{s_gen}
%S&=&\partial (T\ln Z) / \partial T, %\quad
%\\
%\label{n_gen}
%N_B&=&(1/3)\partial (T\ln Z) / \partial \mu_q, %&\quad&
%\\
%\label{e_gen}
%E&=& -PV+TS+3 \mu_q N_B.
%\end{eqnarray}
This is a conventional way of maintaining the thermodynamical consistency in approaches
of the Ginzburg--Landau type as they are applied in high-energy physics, e.g., in the PNJL model.

The exact density matrix $\rho=e^{-\beta {\hat H}}$ of interacting quantum
systems can be constructed using a path integral
approach~\cite{feynm,zamalin}
based on the operator identity
$e^{-\beta \hat{H}}= e^{-\Delta \beta {\hat H}} \cdot
e^{-\Delta \beta {\hat H}}\dots  e^{-\Delta \beta {\hat H}}$,
where the r.h.s. contains $n+1$ identical factors with $\Delta \beta = \beta/(n+1)$.
which allows us to
rewrite\footnote{For the sake of notation convenience, we ascribe superscript $^{(0)}$
to the original variables.}
the integral in Eq.~(\ref{Gq-def}) as follows
\begin{eqnarray}
&&
\sum_{\sigma } \int\limits dr^{(0)}d\mu Q^{(0)}\,
\rho(r^{(0)},Q^{(0)},\sigma ; \{N\};\beta) =
\nonumber\\&=&
\sum_{\sigma} \int\limits  dr^{(0)}d\mu Q^{(0)} dr^{(1)}d\mu Q^{(1)}\dots
dr^{(n)}d\mu Q^{(n)} \, \rho^{(1)}\cdot\rho^{(2)} \, \dots \rho^{(n)}\times
\nonumber\\&\times&
\sum_{P_q} \sum_{P_{ \bar{q}}}\sum_{P_g}(- 1)^{\kappa_{P_q}+ \kappa_{P_{\bar{q}}}}
{\cal S}(\sigma, {\hat P_q}{\hat P_{ \bar{q}}}{\hat P_g} \sigma^\prime)\,
%\times \nonumber\\&&
{\hat P_q} {\hat P_{ \bar{q}}}{\hat P_g}\rho^{(n+1)}\big|_{r^{(n+1)}= r^{(0)}, Q^{(n+1)}= Q^{(0)}, \sigma'=\sigma} =
\nonumber\\
&=&
%&\approx&
\int\limits d\mu Q^{(0)}\int\limits dr^{(0)} dr^{(1)}\dots dr^{(n)}
\tilde{\rho}(r^{(0)},r^{(1)}, \dots r^{(n)};Q^{(0)};\{N\};\beta),
 \label{Grho-pimc}
\end{eqnarray}
where $\{N\} = (N_u,N_d,N_s,\bar{N}_{u},\bar{N}_{d},\bar{N}_{s},N_g)$ is introduced for briefness.
The spin gives rise to the spin part of the density matrix (${\cal
S}$) with exchange effects accounted for by the permutation
operators  $\hat P_q$, $\hat P_{ \bar{q}}$ and $\hat P_g$ acting on the quark, antiquark and
gluon 
%spatial $r^{(n+1)}$, color $Q^{(n+1)}$ 
degrees of freedom  and the spin projections $\sigma'$. The
sum runs over all permutations with parity $\kappa_{P_q}$ and
$\kappa_{P_{ \bar{q}}}$.
In Eq.~(\ref{Grho-pimc})
\begin{eqnarray}
\rho^{(l)}%\equiv
=
\rho\left(r^{(l-1)},Q^{(l-1)};r^{(l)},Q^{(l)};\{N\};\Delta\beta\right) =
\langle r^{(l-1)}|e^{-\Delta \beta {\hat H}}|r^{(l)}\rangle\delta(Q^{(l-1)}-Q^{(l)}),
%\delta_{f^{(l-1)},f^{(l)}},
\label{rho(l)}
\end{eqnarray}
is the off-diagonal element of the density matrix.
Since the color charge is treated classically, we keep only diagonal terms
($\delta(Q^{(l-1)}-Q^{(l)})$) in color degrees of freedom.
Accordingly each quasiparticle is represented by a set of coordinates
$\{r_i^{(0)}, \dots , r_i^{(n)}\}$ (``beads'')
%$\{{\bf r}_i^{(0)}, \dots {\bf r}_i^{(n)}\}$ (``beads'')
and a 8-dimensional color
vector $Q_i^{(0)}$ in the $SU(3)$ group.
Thus, all "beads" of each quasiparticle are characterized by the same spin projection,  flavor and color charge. Notice that masses and coupling
constant in each $\rho^{(l)}$ are the same as those for the original quasiparticles, i.e. these are still defined by the actual temperature $T$.  

The main advantage of decomposition (\ref{Grho-pimc}) is that it
allows us to use perturbation theory to obtain approximation for density matrices $\rho^{(l)}$,  
which is applicable due to smallness of artificially introduced factor $1/(n+1)$. 
%From physical point of view 
This means that
%the effective temperature 
in each $\rho^{(l)}$  the ratio 
% $1/\Delta \beta = (n+1)T$ 
%$\frac{g^2(T,\mu_q) \langle Q_i|Q_j \rangle}{4\pi| r_i- r_j|(n+1)T}$ 
$g^2(T,\mu_q) \langle Q^{(l)}_i|Q^{(l)}_j \rangle /4\pi| r^{(l)}_i- r^{(l)}_j|T(n+1)$ 
can be always made much smaller than one 
%larger then the potential energy per ``bead''
, which allows us to use perturbation theory with respect to the potential.
%From physical point of view 
%this means that the characteristic distance between subsequent ``beads'' %$r_a^{(l-1)}$  and  $r_a^{(l)}$ 
%for each particle in Eq.~(\ref{Grho-pimc}) can be always made  
%This parameter makes the thermal  wavelength $\Delta\lambda_a=\lambda_c\sqrt{2 \pi (m_a/T)/(n+1)}$
%of a bead of type $a$ ($a = qF, \overline{q}, g$),
%smaller then a characteristic scale of variation of the potential energy. 
Each factor in (\ref{Grho-pimc}) should be calculated with the accuracy of order of $1/(n+1)^\theta$ 
with $\theta > 1 $, as in this case the error of the whole product in the limit of large $n$ 
will be equal to zero. %of order $(n+1)/(n+1)^\theta \rightarrow 0$  at 
%Let us now consider approximations to the density matrices $\rho^{l}$.
%An approximation which is suitable for direct PIMC simulations has the following form.
In the limit $(n+1)\longrightarrow \infty$ $\rho^{l}$ can be approximated by a product of two-particle density matrices $\rho_{ij}^{(l)}$ \cite{feynm,zamalin,filinov_ppcf01}. 
Generalizing the electrodynamic plasma results \cite{filinov_ppcf01} to the quark-gluon plasma case,
we write approximate $\tilde{\rho}$
%%%%%%\\
%%%%%%
%%%%%%{\rm
%%%%%%оЕПЕДЕКЮК  Eq.~(\ref{Grho_s})
%%%%%%\\
%%%%%%}
%%%%%%
%
\begin{eqnarray}
&&
\tilde{\rho}(r^{(0)},r^{(1)}, \dots r^{(n)};Q^{(0)}; \{N\};\beta)=
\nonumber\\&=&
 %\sum_{\bar{s},\bar{k}}
 %\frac{C^s_{N_q}}{2^{N_q}} \frac{C^k_{N_{ \bar{q}}}}{2^{N_{\bar{q}}}}
\exp\{-\beta U\}
 \,
 \frac{{\rm per}\,||\widetilde{\underline{\phi}}^{(n),(0)}||_{N_g}}{\tilde{\lambda}_g^{3N_g}}
 \,
%\nonumber\\&&
%\prod\limits_{q=u,d,s}
\frac{{\det}\,||\tilde{\phi}^{(n),(0)}||_{N_q} \,}{\tilde{\lambda}_q^{3N_q}}
\,
%\prod\limits_{\bar{q}=\bar{u},\bar{d},\bar{s}}
\frac{{\det}\,||\tilde{\phi}^{(n),(0)}||_{N_q} \,}{\tilde{\lambda}_{{\bar{q}}}^{3\bar{N}_q}}
\,
\prod\limits_{l=1}^n \prod\limits_{i=1}^N
\phi^{(l)}_{ii} \, , \label{Grho_s}
\end{eqnarray}
%
%%%%%%\\
%%%%%%
%%%%%%{\rm
%%%%%%$r$ МЕ НОПЕДЕКЕМН.
%%%%%%
In Eq.~(\ref{Grho_s}) the effective total color interaction energy 
\begin{eqnarray}
U = 
%\frac{1}{n+1}\sum_{l=1}^{n+1}\tilde{U}^{(l) }=
\frac{1}{n+1}\sum_{l=1}^{n+1} \frac{1}{2}
\sum_{i,j (i\neq j)}^N
\Phi_{ij}(|r_i^{(l-1)}-r_j^{(l-1)}|,|r_i^{(l)}-r_j^{(l)}|, Q_i^{(0)},Q_j^{(0)}).
\label{up}
\end{eqnarray}
is described in terms of the off-diagonal elements of the effective potential 
approximated  by the diagonal ones by means of
$\Delta\beta\Phi^{ij}(x_{ij}^{(l-1)},x_{ij}^{(l)};,Q_i,Q_j)\approx \Delta\beta  \left[\Phi^{ij}(x_{ij}^{(l-1)},x_{ij}^{(l-1)},Q_i,Q_j) + \Phi^{ij}(x_{ij}^{(l)},x_{ij}^{(l)},Q_i,Q_j)\right]/2 
\propto - \ln\left(\rho_{ij}^{(l)}\right)$. Here the the diagonal 
two-particle effective quantum Kelbg potential is 
\begin{eqnarray}
%\Phi^{pt}( r_{},r_{i},  r_{j},  r_{j},Q_p,Q_t,\Delta\beta) =
\Phi^{ij}( x_{ij}^{(l)},x_{ij}^{(l)},Q_i,Q_j,\Delta\beta) %=
%\Phi^{pt}(x_{pt},Q_p,Q_t,\Delta\beta)
%\, \nonumber\\
= \frac{g^2(T,\mu_q)\,\langle Q_i|Q_j \rangle}{4 \pi %\Delta\lambda_{ij} 
| r_{i}^{(l)}-r_{j}^{(l)}|} \,\left\{1-e^{-(x_{ij}^{(l)})^2} +
\sqrt{\pi} x_{ij}^{(l)} \left[1-{\rm erf}(x_{ij}^{(l)})\right] \right\},
%\label{kelbg-d}
\end{eqnarray}
with $x_{ij}^{(l)}=| r_{i}^{(l)}-r_{j}^{(l)}|/\Delta\lambda_{ij}$,  $\Delta\lambda_{ij}=\sqrt{2\pi\Delta\beta /m_{ij}}$, $m_{ij}=m_{i}m_{j}/(m_{i}+m_{j})$ 
Other quantities in  Eq.~(\ref{Grho_s}) are defined as follows: %$N=N_q+\bar{N}_{q}+N_g$,
$\widetilde{\lambda}^3_a=\lambda^3_a\sqrt{0.5\pi/(\beta m)^5}$ with
  $\lambda_a=\sqrt{2 \pi \beta / m_a}$
 being a thermal  wavelength of an $a$ type quasiparticle ($a=q, \bar{q},g$).
%$C^s_{N_a}=N_a!/[s!(N_a-s)!]$,
The antisymmetrization and symmetrization takes into account quantum statistics and results 
in appearing permanent for gluons and determinants for quarks/antiquarks. 

%Functions $\phi^{(l)}_{ii}\equiv K_2(z_i^{(l)})/(z^{(l)})^2_i$ ($z_i^{(l)}=\Delta\beta m_i(T,\mu_q)\sqrt{1+2\pi %\Delta\beta \left|\xi^{(l)}_i\right|^2 /m_i(T,\mu_q)}$ 
Functions $\phi^{(l)}_{ii}\equiv K_2(z_i^{(l)})/(z^{(l)})^2_i$ ($z_i^{(l)}=\Delta\beta m_i(T,\mu_q)\sqrt{1+
  \left|\xi^{(l)}_i\right|^2 /\Delta\beta^2}$ 
%(proportional to off-diagonal matrix elements of relativistic one particle density matrices) 
are defined by modified Bessel functions.    
%$\phi^l_{pp}\equiv \exp\left[-\pi\left|\xi^{(l)}_p\right|^2\right]$  
Gluon matrix elements are 
$\widetilde{\underline{\phi}}_{i,j}^{(n),(0)}=K_2(z_{i,j}^{(n),(0)})/(z_{i,j}^{(n),(0)})^2\delta_\epsilon(Q^{(0)}_i-Q^{(0)}_j)$, 
 while quark and antiquark  matrix elements 
$\tilde{\phi}_{i,j}^{(n),(0)}=K_2(z_{i,j}^{(n),(0)})/(z_{i,j}^{(n),(0)})^2\delta_\epsilon(Q^{(0)}_i-Q^{(0)}_j)\delta_{f_i,f_j}\delta_{\sigma_i,\sigma_j}$
 depend additionally on spin variables $\sigma_i$ and flavor index $f_i$ of the particle, which  can take values ``up'', ``down'' and ``strange'', $\delta_{\sigma_i,\sigma_j}$ and $\delta_{f_i,f_j}$ are the Kronecker's deltas, while $\delta_\epsilon$ is the delta-like function of color vectors. These functions allow  to exclude the Pauli blocking for particles with different  spins, flavors and colors.  Here arguments of modified Bessel functions are 
$z_{i,j}^{(n),(0)}=\Delta\beta m_i(T,\mu_q) \sqrt{1+\left|r_{i}^{(0)}-r_{j}^{(n)}
\right|^2/\Delta\beta^2}$. 
%$z_{i,j}^{(n),(0)}=\Delta\beta m_i(T,\mu_q) \sqrt{1+2\pi\Delta\beta\left|r_{i}^{(0)}-r_{j}^{(n)}
%%+ y_{i}^{(n)}
%\right|^2/\Delta\lambda_i^2 m_i(T,\mu_q)}$. 
%\right|^2/(\Delta\lambda^2_{i}\Delta\beta m_i(T,\mu_q))}$. 
%$\tilde{\phi}_{to}^{n,0}=\exp \left(-\pi
%\left|(r_{t}^{(0)}-r_{o}^{(0)})+ y_{t}^{(n)}\right|^2/\Delta\lambda_{a}^2\right)
%\delta_\epsilon(Q_t-Q_o)$
%Here $t$ and $o$ are particle's indexes, 
The coordinates of the quasiparticle ''beads''  $r_{i}^{(l)} = r_{i}^{(0)}+y_{i}^{(l)}$, ($l>0$) are expressed in terms of $r_{i}^{(0)}$ and 
%The result can be rewritten in terms of
%coordinates $r$ which depend on the
vectors between
neighboring beads of an $i$ particle, defined as
%$y_i^{(l)}=\Delta\lambda_i\sum_{k=1}^{l}\xi_i^{(k)}$. Here 
$y_i^{(l)}= \sum_{k=1}^{l}\xi_i^{(k)}$,  %Here 
%$\Delta\lambda_i=\sqrt{2 \pi \Delta\beta / m_i}$, 
while  $\xi_{i}^{(1)}, \dots , \xi_{i}^{(n)}$ are 
%($y_{i}^{(1)}, \dots , y_{i}^{(n)}$) and
vector variable of integration  in Eq.~(\ref{Grho-pimc}).  
%; r^{(1)};r^{(2)}; \dots;r^{(n)}\}\equiv
%[r_{a}^{(0)}; r_{a}^{(0)}+y_{a}^{(1)};r_{a}^{(0)}+_{a}y^{(2)}; \dots;r_{i}^{(0)}+y^{(n)}\}$,
%with
%$y^{(l)}\equiv [y_q^{(l)},y_{\bar{q}}^{(l)},y_g^{(l)}]$,
%, with $a=q, \bar{q}, g $ specifying the  type of particles,
%Notice that 
%the indices $\bar{s}$ and $\bar{k}$ in $\det\,||\tilde{\phi}^{n,0}||_{\bar{s}}$and  
%$\det\,||\tilde{\phi}^{n,0}||_{\bar{k}}$  denoted that 
%matrices $||\tilde{\phi}^{(n),(0)}||$ can be transformed to the form having nonzero blocks 
%related to quark and antiquark quasiparticles with the same color,  flavour and spin projections. 

The main contribution to the partition function comes from
configurations in which  ``size '' of the quasiparticles cloud of 'beads' is of
order of Compton wavelength $\lambda_C=1/m(T,\mu_q)$ providing spatial quasiparticle localization. 
So this path integral representation takes into account quantum position uncertainty of quasiparticles. 
In limit of large mass spatial quasiparticle localization can be made much smaller than average 
interparticle distance. This allow analytical integration over 'beads' position by method of 
steepest decent. As result the partition function is reduced to its classical limit with point strongly 
interacting color quasiparticles.  
%Determinants are flavor sensitive, while quasiparticle interaction 
%$U$ does not depend on flavor indices.  
%The density matrix (\ref{Grho_s}) has been transformed to a form which does not
%contain an explicit sum over permutations
% and thus no sum of terms with alternating sign (in the case of quarks and antiquarks).

In the limit of $n\rightarrow\infty$ functions $\phi^{(l)}_{ii}$ describe the {\em new relativistic measure} of developed color path integrals. This measure is created by {\em relativistic operator of kinetic energy} $K=\sqrt{p^2+m^2(T,\mu_q)}$. Let us note that in the limit of large particle mass the relativistic measure coincide with the Gaussian one used in Feynman and Wiener path integrals. 

%For the density matrices $\rho^{(l)}$ of the ``bead'' the effective temperature
%$(n+1)T$ is already higher then the quasiparticle mass. This is the reason why we %kept the relativistic
%kinematics in the Hamiltonian of Eq. (\ref{Coulomb}).

%----------------------------
\section{Simulations of QGP}\label{s:model}

\begin{figure}[htb]%\label{fig:cor}
\vspace{0cm} \hspace{0.0cm}
\includegraphics[width=7.8cm,clip=true]{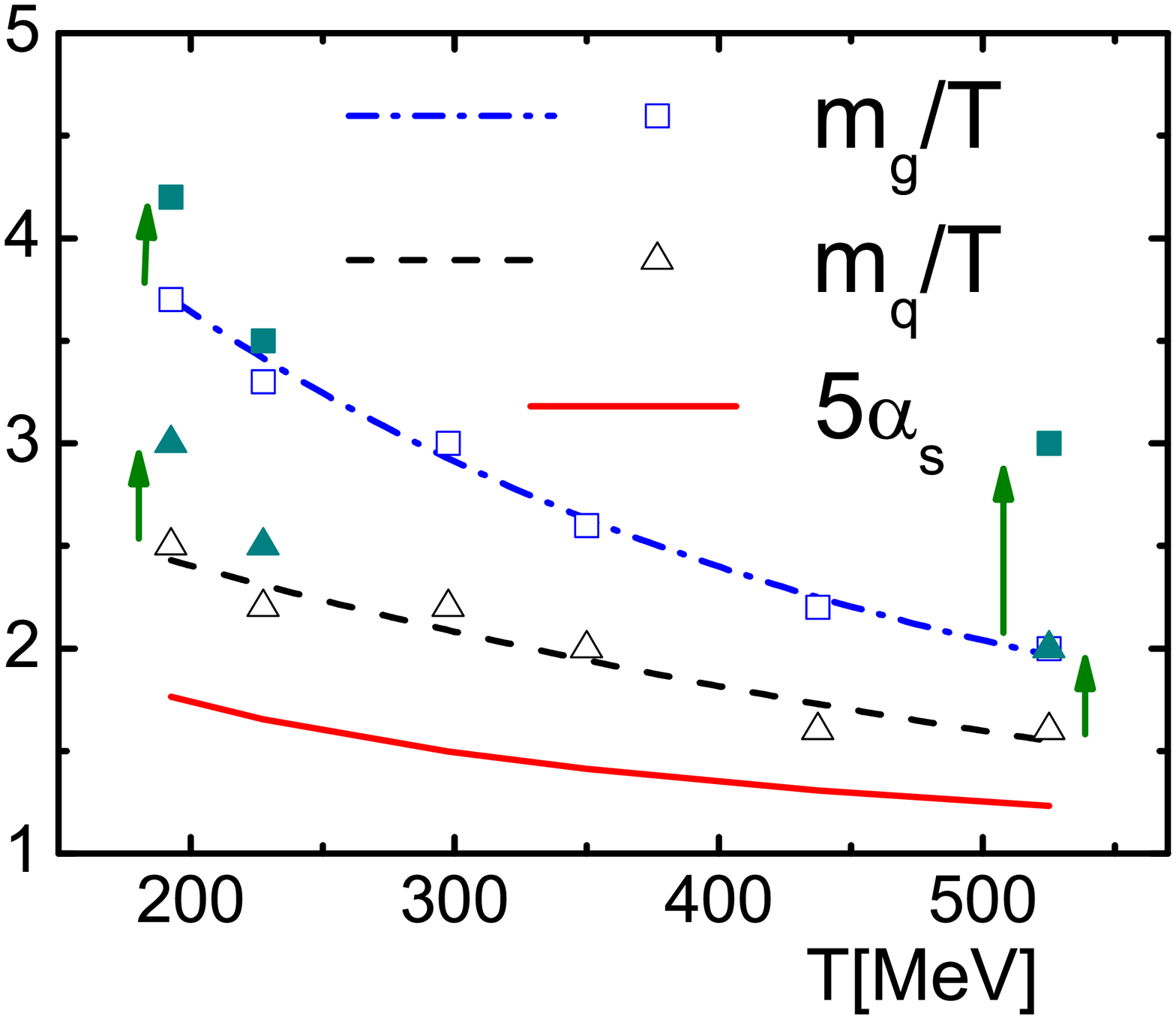}
\includegraphics[width=7.8cm,clip=true]{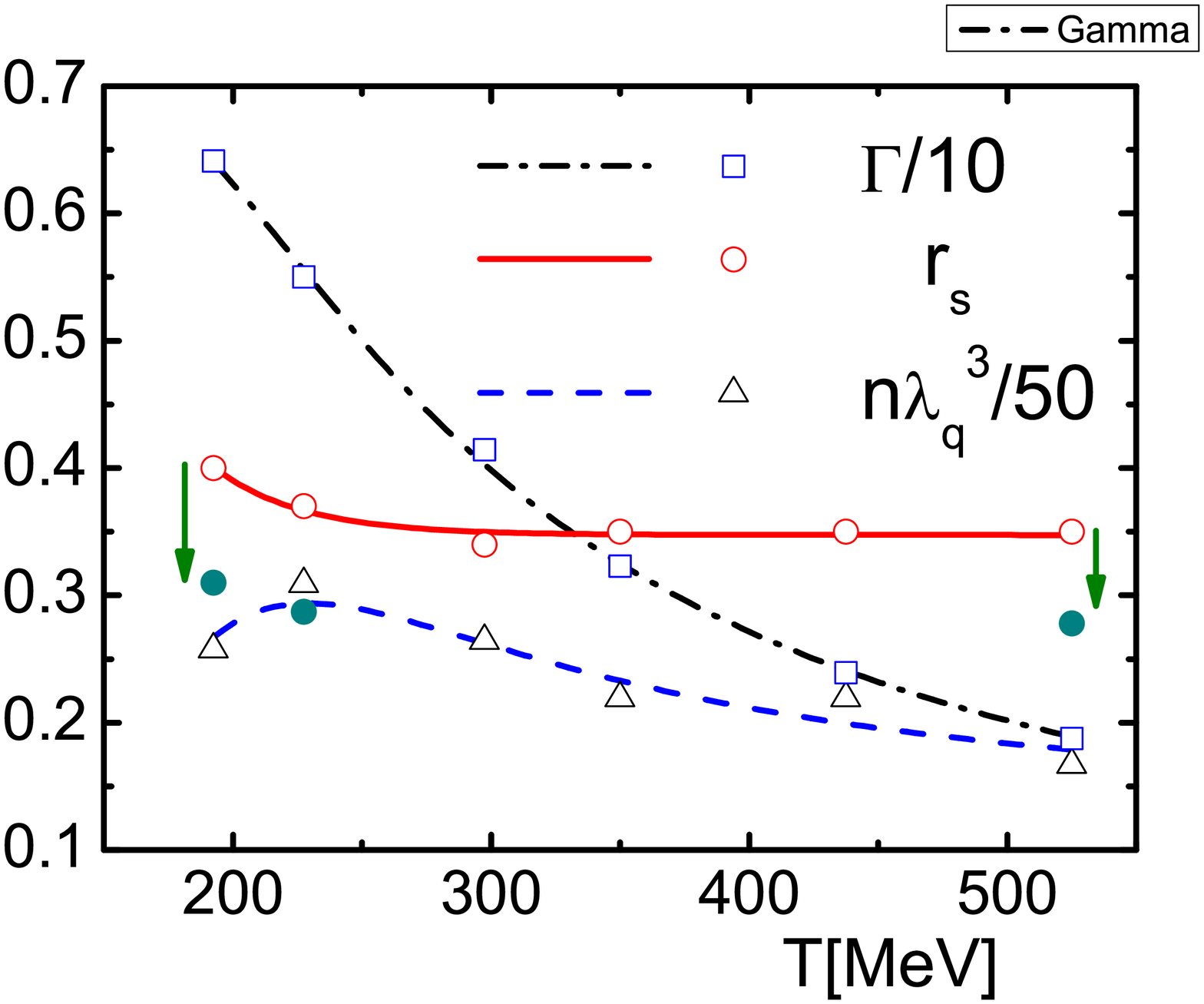}
\caption{
Left panel:Temperature dependences of the model input quantities:
the coupling constant $\alpha_s$ (scaled by 5) and quasiparticle
mass-to-temperature ratio ($m_q=m_{\bar{q}}$). 
Right panel: Temperature dependences of the calculated interparticle average distance $r_s$ (Wigner-Seitz radius), 
the quark quasiparticles  degeneracy parameter $\chi$
and the plasma coupling parameter $\Gamma$. % [see Eq. (\ref{Gamma})].
%The degeneracy parameters for different species does not coincide, since the
%quasiparticle masses are different.
Lines are smooth interpolation between the CPIMC points. 
To ilustrate sensibility of this model to input data the green solid symbols and arrows 
show changes of the CPIMC average interparticle distance (right panel) related to 
the quasiparticle mass variation on left panel.   
}
\label{fig:alfrs}
\end {figure}

To test the developed approach we consider the QGP at zero baryon density with equal flavor fractions of quarks and antiqiarks
($N_u=N_d=N_s=N/3=\bar{N}_u=\bar{N}_d=\bar{N}_s=\bar{N}/3$). 
Ideally the parameters of the model should be deduced from the QCD lattice data. However, presently this task is still quite ambiguous. Therefore, in the present simulations we take only a possible set of parameters.
The phenomenologic QCD estimations \cite{Prosperi} of coupling constant,
i.e. $\alpha_s = g^2/(4\pi)$, used in the simulations is displayed in the left panel of Fig.~\ref{fig:alfrs}.
%Following  \cite{LiaoShuryak,shuryak1},
%The parametrization of the quasiparticle mass is taken in the form
%%
%\begin{eqnarray}
%\label{mass}
%m(T)/T_c=0.9/(T/T_c-1)+3.45+0.4T/T_c
%\end{eqnarray}
%%
%where $T_c=175$ MeV is the critical temperature. This parametrization fits the quark mass at two values
%of temperature obtained in the lattice calculations \cite{Lattice02}. According to \cite{Lattice02} the masses
%are quite large: $m_q/T \simeq 4$  and $m_g/T \simeq 3.5$. These are essentially larger than masses
%required for quasiparticle fits \cite{Peshier96,Ivanov05} of the lattice  thermodynamic properties
%of the QGP:  $m_q/T \simeq 1\div 2$  and $m_g/T \simeq 1.5\div 3$.
%Moreover, the pole quark mass $m_q/T \simeq 0.8$ was reported
%in recent work \cite{Karsch09a}, as deduced from lattice calculations. Nevertheless, in spite of the fact
%that it obviously produces too high masses, we use parametrization
%(\ref{mass}) in order to be compatible with the input of classical MD
%of   \cite{shuryak1}.
The $T$-dependence of quasiparticle  mass used in this work is also presented in Fig.~\ref{fig:alfrs}
(left panel). The right panel presents the calculated in grand canonical ensemble
the temperature dependences of the
interparticle average distance $r_s$ (Wigner-Seitz radius $r_s^3=(3/4\pi n)$,
n is the density of all quasi particles). The quark quasiparticles
degeneracy parameter $\chi$ and the plasma coupling parameter $\Gamma$ are defined as:
%[see Eq. (\ref{Gamma})]:
$%\begin{eqnarray}
%\label{Gamma}
\chi=n_q\lambda^3_q,   \quad \quad 
\Gamma = \frac{ \overline{q_2} g^2}{4\pi r_s T},
$%\end{eqnarray}
where
%$r_s$ is the the Wigner-Seitz radius, defined such that $4\pi r_s^3/3 = n $, and
$\overline{q_2}$ the quadratic Casimir value
averaged over quarks, antiquarks and gluons,
$\overline{q_2}=N_c^2-1$ is a good estimate for this quantity.
The plasma coupling parameter is a measure of ratio of the average potential
to the average kinetic energy. It turns out that $\Gamma$ is larger the
unity which indicates  that the QGP is a strongly coupled Coulomb liquid
rather than a gas.
In the studied temperature range, $190 \text{ MeV} < T< 600 \text{ MeV}$, the QGP  is, in fact,
quantum degenerate, since
%i.e. spin-color statistics plays a significant role.
the degeneracy parameter
$\chi = n_q\lambda_q^3$
(where the thermal wave length $\lambda_q$ was defined in the
previous sect.) 
varies from $\sim 20$ to $\sim 8$, see Fig.~\ref{fig:alfrs} (right panel).
The degeneracy parameters for different species does not coincide, since the
quasiparticle masses are different.
\begin{figure}[htb]%\label{fig:cor}
\vspace{0cm} \hspace{0.0cm}
\includegraphics[width=7.8cm,clip=true]{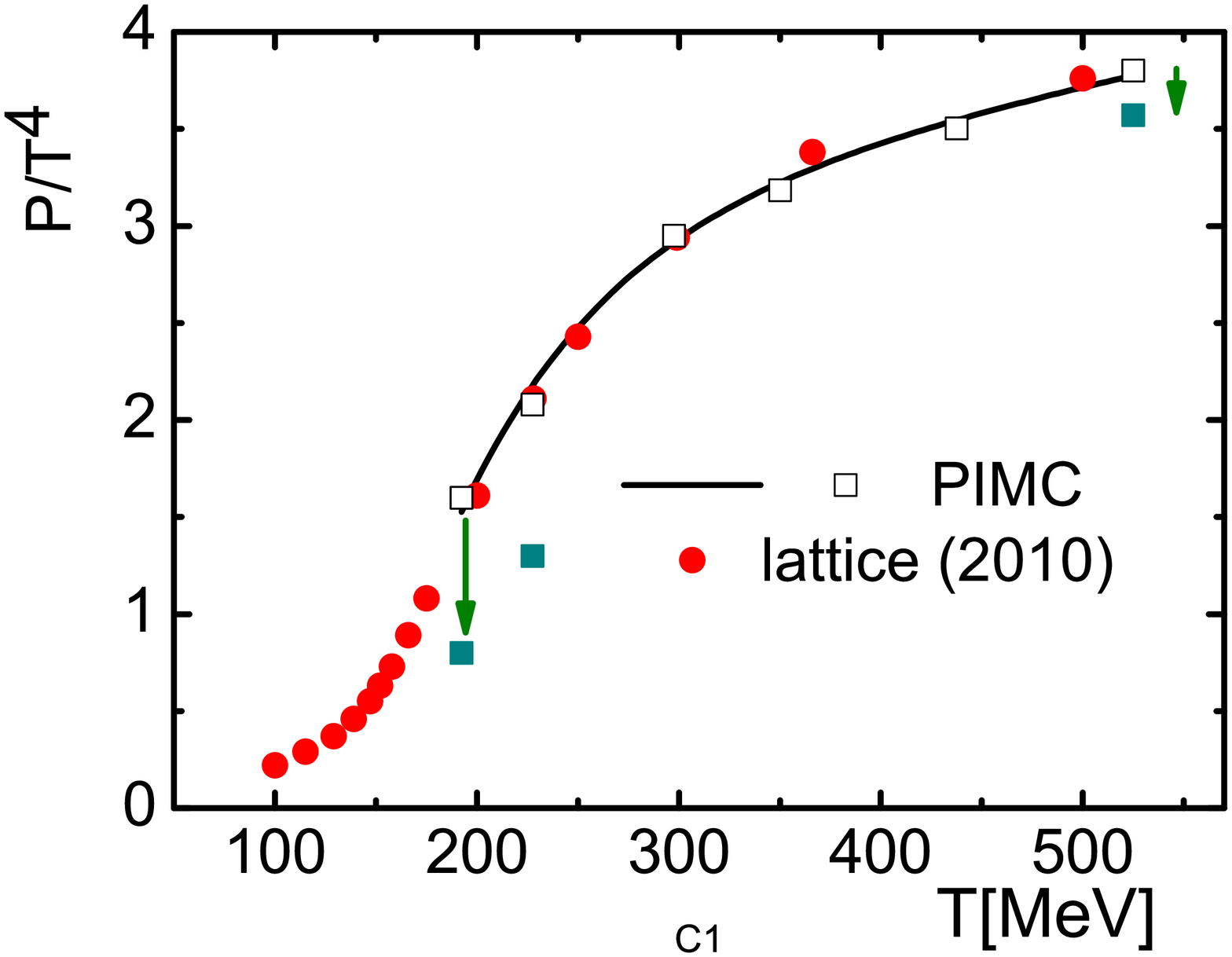}
\includegraphics[width=7.8cm,clip=true]{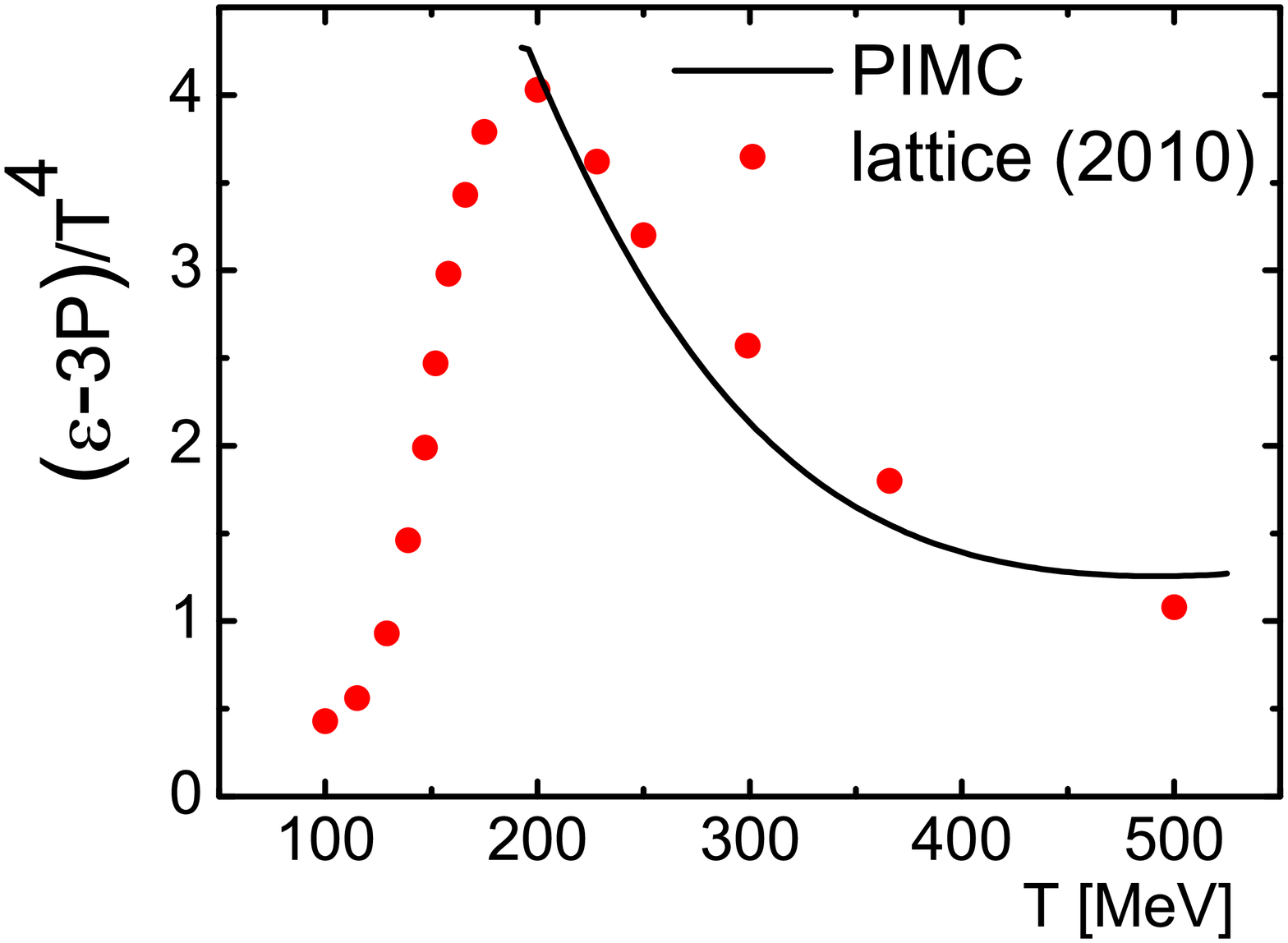}
\caption{
%Left panel: Temperature dependence of the model input
%  quantities, coupling constant $\alpha_s$ (scaled by 10) and mass-to-temperature ratio,
%the plasma coupling parameter $\Gamma$ [see Eq. (\ref{Gamma})] and the
%degeneracy parameter $\chi$. The $\chi$ parameters for different
%species are equal, since their masses and densities are assumed to be
%equal.
%
%Right panel:
Equation of state (left panel) of the QGP from CPIMC
simulations (open squares). 
The solid line is a smooth interpolation between the CPIMC points. 
The solid symbols illustrate the changes in equation of state for presented by  Fig.~\ref{fig:alfrs}  
quasiparticle mass variation.   
Trace anomaly (right panel) of the QGP from CPIMC simulations is presented by solid line. 
%compared to lattice data of \cite{Fodor09,Csikor:2004ik}.
Both CPIMC results are compared  to lattice data of  \cite{Fodor09,Csikor:2004ik}
}
\label{fig:EOS}
\end {figure}
As follows from analysis of Fig.~\ref{fig:alfrs} and Fig.~\ref{fig:EOS} 
this model is highly  sensitive to quasiparticle mass variations. This 
allow to fit lattice EOS and to choose optimal values of gluon and quark masses  
%was used to optimize the parameters of the model 
in order to proceed in predictions of other properties concerning the internal structure and in the future also non-equilibrium dynamics of the QGP. 
High sensitivity of EOS to mass variation adds to background of  this procedure. 
Let us also note that increase of quasiparticles masses reduces the influence of quantum effects, 
so high sensitivity of EOS to masses proved the importance of quantum effects at these 
temperatures.  
%%
%\begin{figure}[htb]%\label{fig:cor}
%\vspace{0cm} \hspace{0.0cm}
%\includegraphics[width=7.9cm,clip=true]{FGST3.eps}
%\includegraphics[width=7.9cm,clip=true]{FGTrAn.eps}
%\caption{
%Entropy (left panel) and trace anomaly (right panel) of the QGP from CPIMC
%simulations (solid line) compared 
%to lattice data of \cite{Fodor09,Csikor:2004ik}.
%}
%\label{fig:EOS1}
%\end {figure}
%%
Figure \ref{fig:EOS} presents also the %the entropy ($S/T^3$) and
trace anomaly ($(\varepsilon-3P)/T^4$)  %of the QGP computed by the CPIMC method. These function is 
calculated as related combination of derivatives of equation of state. 
%accordingly to Eqs. (\ref{p_gen}). 
In order to avoid the numeric noise, the derivatives of a smooth interpolation between the
CPIMC points of equation of state (Fig. \ref{fig:EOS}) were taken. The CPIMC results are  compared
to lattice data of  \cite{%Lattice09,
Fodor09,Csikor:2004ik}. It is not surprising that agreement with the lattice data for trace anomaly 
is also good, since it is a direct consequence of the good reproduction of the pressure.

\begin{figure}[htb]%\label{fig:PDFC}
\vspace{0cm} \hspace{0.0cm}
\includegraphics[width=7.7cm,clip=true]{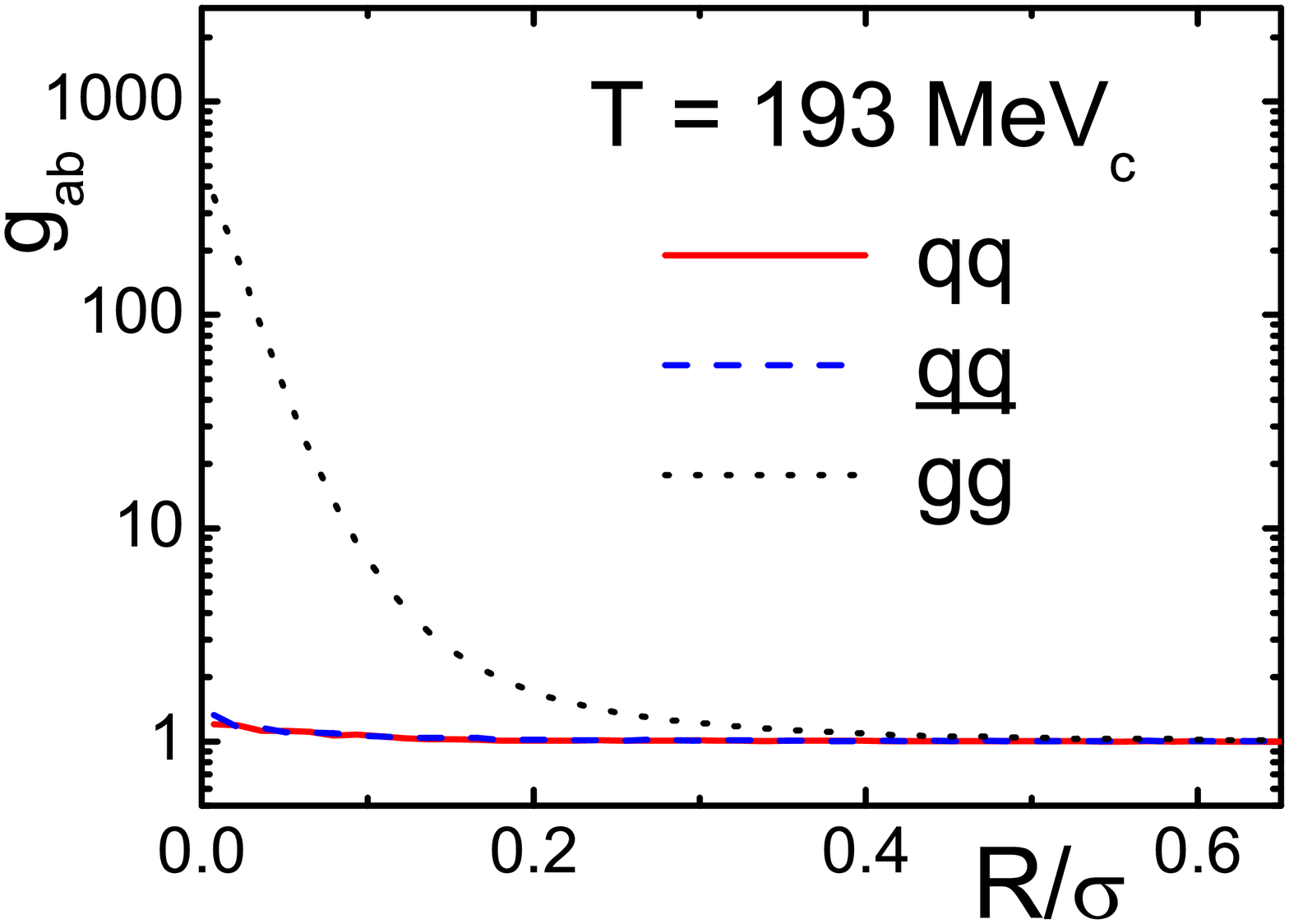}
\includegraphics[width=7.7cm,clip=true]{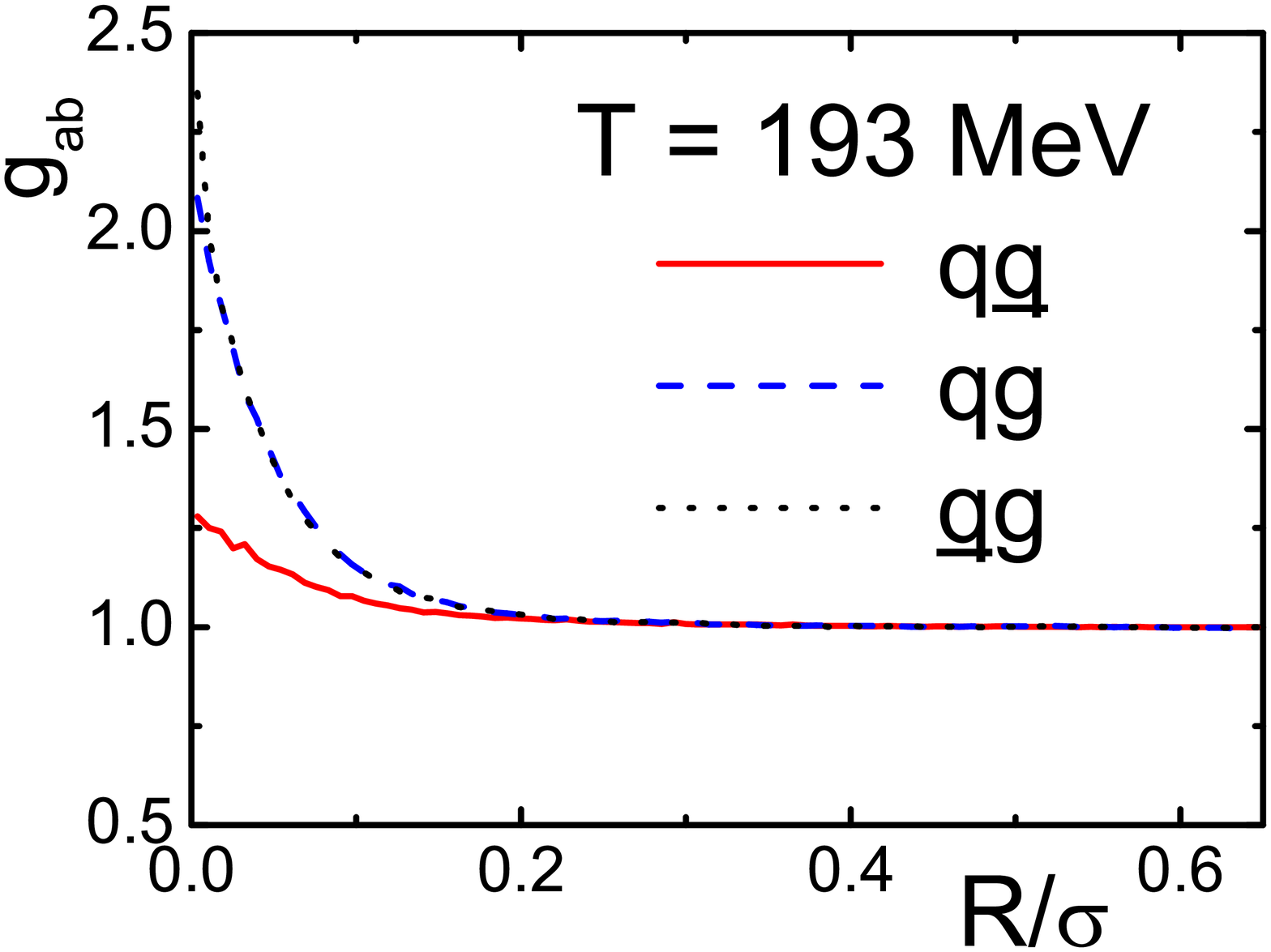}
\caption{Pair distribution functions
%(upper panels) and color pair distribution functions (lower panels)
of identical (left panel) and different (right panel) quasiparticles  at
temperature 
%$T=525 \text{ MeV}$ (upper panels) and 
$T=193 \text{ MeV}$. % (lower panels).
The distance is measured in units of $\sigma = 1.1$ fm.
}
\label{fig:PDFC}
\end{figure}

Let us now consider the spatial arrangement of the quasiparticles in the QGP by studying the pair distribution functions (PDF) $g_{ab}(R)$. They give the probability density to find a pair of particles of types $a$ and $b$ at a certain distance $R=|R_1-R_2|$ from each other and are defined, for example, in canonical ensemble as
%%----------------------------------------------------------------------------------------------------------------
\begin{eqnarray}\label{g-def}
%g_{ab}(|R_1,R_2|) &=&
g_{ab}(|R_1-R_2|)=
%\nonumber\\
%\frac{1}{{\tilde Z}}
\frac{1}{Z\left(N,V,\beta\right) \text{,}}
\sum_{\sigma}\int \limits_V 
dr\; dQ ,\delta(R_1-r^a_1)\delta(R_2-r^b_2)\;\rho(r,Q, \sigma; N_u,N_d,N_s,\bar{N}_{u},\bar{N}_{d},\bar{N}_{s},N_g; \beta),
%dr dQ\,\delta(R_1-r^a_1)\delta(R_2-r^b_2)\rho(r,Q, \sigma ;\beta).
%\\
%{\tilde Z} &= &
%Z(N_q,N_{ \bar{q}},N_g,V;\beta) N_q!N_{ \bar{q}}!N_g!.
\end{eqnarray}
The PDF depend only on the difference of coordinates because of the translational invariance of the system.
In a non-interacting classical system,
$g_{ab}(R)\equiv 1$, whereas interactions and  quantum statistics result in
a re-distribution of the particles.
At temperatures %$T=525 \text{ MeV}$ and 
$T=193 \text{ MeV}$  the PDF averaged over the quasiparticle spin, colors and flavors are shown in %Fig.~2,
the panels of Fig.~\ref{fig:PDFC}.
%
%PDF's of identical particles are presented in the upper left panel of Fig.~\ref{fig:PDFC}.
%

At distances, $R/\sigma \geq r_s (r_s\approx 0.35)$ all PDF are practically equal to unity (Fig. \ref{fig:PDFC}) like in ideal gas due to the 'Debye' screening of the color Coulomb interaction. 
%, approaching unity, i.e. the ideal-gas limit.
A drastic difference in the behavior of the
PDF of  quarks and gluons (the anti-quark PDF is identical to the quark PDF) occurs at distances $R/\sigma \le r_s$. Very well pronounced is the short-range structure of nearest neighbors. Here the gluon-gluon and gluon-quark PDF increases monotonically when the distance goes to zero.  
Behavior of these PDF %of the PDF at small distances
is a clear manifestation of an effective pair attraction.
%of quarks and antiquarks as well as quarks (antiquarks) and gluons.
Attraction suggests that the color vectors of nearest
neighbor quasiparticles are anti-parallel.
The QGP lowers its total energy by minimizing the color 
Coulomb interaction energy via a spontaneous ``anti-ferromagnetic'' ordering of color vectors. % of gluons. 
From physical point of view this means formation of the gluon-gluon and gluon-(anti)quark clusters, which are more or less  uniformly distributed in space. % spatial distribution.  
The last conclusion comes from the fact that 
the (anti)quark-(anti)quarks PDF are close to unity.
%This gives rise to a clustering 
%of gluons which is accompanied by a tendency
%of clustering of quark pairs with anti-parallel spins. We also observe clusters
%of quarks, antiquarks and gluons.  
Oscillations of the PDF at very small distances of order
$R\le 0.1 \sigma$ are related to Monte Carlo statistical error, as probability of  quasiparticles being at short distances quickly decreases.
%exhibits a broad maximum. 

The main physical reason for arising the well separated from each other the  gluon-gluon and gluon-(anti)quark) quasiparticle clusters is the consequence of the great difference in values of quadratic Casimir invariants $q_2=<Q_i^{(l)}|Q_i^{(l)}>$ for gluon-gluon $q_2=24$, (anti)quark-(anti)quark  $q_2=4$ and gluon-(anti)quark $q_2=\sqrt{24*4} \approx 10$ in color Kelbg (Coulomb) potentials 
%describing quasiparticles interaction (\ref{kelbg-d}). 
 of quasiparticles interactions . 

The second physical reason of the PDF difference is spatial quantum uncertainty and different property of Bose and Fermi statistics of gluon and (anti)quark quasiparticles. Fermi statistics results in effective quark-quark and aniquark-antiquark repulsion, while Bose one results to effective qluon-qluon attraction.  More strong interaction of the QGP quasiparticles  with gluons than with quarks is also consequence of the gluon better localization in comparison with quarks. As we mentioned before uncertainity in particle localization is defined by the ratio $1/m$. Localization is better for heavier gluon quasipartcles. To verify the relevance of all these trends, a more refined color, flavor, and  spin-resolved analysis of the PDF is necessary, together with simulations
in a broader range of temperatures which are presently in progress.

Details of our path integral Monte-Carlo simulations have been discussed elsewhere in a variety of papers and review articles, see,
e.g. \cite{zamalin,rinton} and references therein. The main idea of the simulations consists in constructing a Markov chain of quasiparticle configurations. 
% (including all beads).
Additionally to the case of electrodynamic plasmas here  we  randomly sample the color quasiparticle variables $Q$ according to the group SU(3) Haar measure. 
% with the quadratic and cubic Casimir conditions until convergence is achieved. Next important difference with electrodynamic plasma consist in using here for path integral Monte Carlo simulations the relativistic measure created by relativistic kinetic energy operator instead of simulation with Gaussian one arising from related non relativistic operator used in Feynman or Wiener path integrals. As usually for simulations we use a cubic simulation box with periodic boundary conditions. The number of particles was taken as $N=N_q+N_{ \bar{q}}+N_g=42+42+42=126$, and the number of "beads", $n=20$. 

%%%%%%%%%%%%%%
%----------------------------
\section{Conclusion}\label{s:discussion}

%Experimental data on the quark-gluon plasma and the hadronization transition give rise
%to numerous challenges to the theory.  %, see, e.g. \cite{shuryak08,biro08} and references therein.
%Of particular interest is the question why the
%quark-gluon matter behaves as an almost perfect fluid rather than as a perfect gas, as it
%could be expected from the asymptotic freedom.
Color quantum Monte Carlo simulations %\cite{Filinov:2009pimc} and quantum molecular dynamics
based on the quasiparticle model of the QGP
%with color Coulomb interactions help us to answer this question.
are able to reproduce the lattice equation of state and other thermodynamic functions (even near the critical temperature) and also
yield valuable insight into the internal structure of the QGP.
Our results indicate that the QGP reveals quasiparticle clustering and liquid-like (rather than gas-like) properties even at the
highest considered temperature. 
% of $525 \text{MeV}$.
%At temperatures just above $T_c$ we have found that
%bound quark-antiquark states still survive. These states are bound by effective string-like forces.
%Quantum effects turned out to be of prime importance in these  simulations.

%Our simulation results indicate that the ratio of the potential to
% kinetic energy is of the order of $1\dots 3$, depending on the temperature.. This certainly corresponds to a liquid-like rather than a gas-like  behavior.

%In ~\cite{Filinov:2009pimc} we have shown that the PIMC method captures main trends of the equation of state (even near the critical temperature) and may also yield valuable insight into the internal structure of the QGP, in particular into the pair correlation functions.
%Our PIMC simulations also allow for a selfconsistent analysis of cluster and bound state formation in the QGP.
%Similar questions have been successfully studied before
%in dense astrophysical plasmas \cite{filinov_jetpl00} and electron-hole plasmas in semiconductors \cite{filinov_jpa03}. In fact, first indications
%for clustering in the QGP have been observed and will be studied in more detail in the future. ???

For our simulations we have introduced new {\em relativistic path integral measure} and have developed procedure of sampling color quasiparticle variables according to {\em the group SU(3) Haar measure} with the quadratic and cubic Casimir conditions. 
Our analysis is still simplified and incomplete. It is still confined only to the case of zero
baryon chemical potential. The input of the model also requires refinement.
Work on these problems is in progress.
%We have also performed first simulations of dynamic properties of the QGP  based on %quantum Wigner dynamics.
%

%----------------------------
%\section*{Acknowledgements}
We acknowledge stimulating discussions with Prof. P.~Levai, D.~Blaschke, R. Bock, and H.~Stoecker. 
%Particular we are very grateful to Prof. Yu.B.~Ivanov for fruitful discussions, important remarks and useful help. 

%----------------------------

\bibliographystyle {apsrev}

\begin{thebibliography}{90}
%%
%%1
\bibitem{shuryak08}
%Physics of Strongly coupled Quark-Gluon Plasma
E. Shuryak, Prog. Part. Nucl. Phys. {\bf 62}, 48 (2009).
%E. Shuryak, Prog. Part. Nucl. Phys. {\bf 53}, 273 (2004)
%arXiv:0807.3033v2 [hep-ph]

%3
\bibitem{Fodor09}
%The Phase diagram of quantum chromodynamics.
Z. Fodor and S. D. Katz, arXiv:0908.3341 [hep-ph].
%4

\bibitem{Csikor:2004ik}
%Szabolcs Borsanyi, Gergely Endrodi, Zoltan Fodor, Antal Jakovac,
%Sandor D. Katz, Stefan Krieg, Claudia Ratti, Kalman K. Szabo,
S. Borsanyi, G. Endrodi, Z. Fodor, A. Jakovac, S. D. Katz, S. Krieg, C. Ratti, K. K. Szabo,
%The QCD equation of state with dynamical quarks
JHEP 1011:077,2010

\bibitem{Levai}
P.~Hartmann, Z.~Donko, P.~Levai, G.J.~ Kalman, 
%Molecular dynamics simulation of strongly coupled QCD plasmas
J. Phys. {\bf A42}, 214029 (2006); Nucl. Phys. {\bf A774}, 881-884,  (2006); 
%arXiv: nucl-th/0601017 


%5
\bibitem{LM105}
D. F.~Litim and C.~Manuel,
%Mean feald dynamics in non-Abelian plasmas for classical transport theory,
Phys. Rev. Lett. {\bf 82}, 4981 (1999); % [hep-ph/9902430];
%\bibitem{LM106}
%D.F.~Litim, C.~Manuel,
%Effective transprot equations for non-Abelin plasmas,
Nucl.Phys. B {\bf 562}, 237 (1999); % [hep-ph/9906210];
%\bibitem{LM107}
%D.F.~Litim, C.~Manuel,
%Fluctuations from dissipation in a hot non-Abelin plasmas,
Phys. Rev. D {\bf 61}, 125004 (2000); % [hep-ph/9910348];
%\bibitem{LM108}
%D.F.~Litim, C.~Manuel,
%Deriving effective transport equationfor non-Abelin plasmas,
%hep-ph/00033029910348;
%\bibitem{LM109}
%D.F.~Litim, C.~Manuel,
%Semi-classical transport theory for non-Abelin plasmas,
Phys. Rep. {\bf 364}, 451 (2002), 

\bibitem{LM109}
M.~Laine and C.~Manuel,
%``A Remark on nonAbelian classical kinetic theory,''
Phys.\ Rev.\ D {\bf 65}, 077902 (2002), 
%> [arXiv:hep-ph/0111113].
%
\bibitem{LM110}
%the color charges quantum mechanically, see please
C.~Manuel and S.~Mrowczynski,
%``Local equilibrium of the quark gluon plasma,''
Phys.\ Rev.\ D {\bf 68}, 094010 (2003), 
% [arXiv:hep-ph/0306209].
%as with this approach, one can match the operators that the classical 
%charges cannot
%
\bibitem{LM111}
P.~F.~Kelly, Q.~Liu, C.~Lucchesi and C.~Manuel,
%``Classical transport theory and hard thermal loops in the quark - gluon
%plasma,''
Phys.\ Rev.\ D {\bf 50}, 4209 (1994)
%[arXiv:hep-ph/9406285].
%%CITATION = PHRVA,D50,4209;%
%
%VOLUME 72, NUMBER 22 PHYSICAL REVIEW LETTERS 30 MAY 1994
%Deriving the Hard Thermal Loops of QeD from Classical Transport Theory
%P. F. Kelly,╥ Q. Liu,t C. Lucchesi.l and C. Manuelt
%Center for Theoretical Physics, Laboratory for Nuclear Science, and Department of %Physics,
%Massachusetts Institute of Technology, Cambridge, Massachusetts 02199
%(Received 30 March 1994)
%Classical transport theory is employed to analyze the hot quark-gluon plasma at %the leading order in

%6
\bibitem{Bleicher99}
%Statistical mechanics of semiclassical colored objects.
M. Hofmann, %{\em et al.}, %
M. Bleicher, S. Scherer, {\em et al.},   %L. Neise, H. Stoecker, and W. Greiner,
 Phys. Lett. B {\bf 478}, 161 (2000). % [nucl-th/9908030].
%7
% classical color Coulomb model
%\bibitem{Feinberg} I. Roizen, E. Feinberg, O. Chernavskaja, Phys. Usp.  \textbf{47}, 427 (2004).
%
% classical color Coulomb model
\bibitem{shuryak1} B. A. Gelman, E. V. Shuryak, and I. Zahed, Phys. Rev. C \textbf{74}, 044908 (2006);
%ibid.
\textbf{74}, 044909 (2006).
%
%8
\bibitem{Zahed}
S. Cho and I. Zahed,
Phys. Rev. C {\bf 79}, 044911 (2009); % [arXiv:0812.1736 (nucl-th)];
 Phys. Rev. C {\bf 80}, 014906 (2009); % [arXiv:0812.1741 (nucl-th)];
arXiv:0910.2666 [nucl-th];
arXiv:0910.1548 [nucl-th];
arXiv:0909.4725 [nucl-th];
K. Dusling and I. Zahed,  Nucl. Phys. A {\bf 833}, 172 (2010). % [arXiv:0904.0169 (nucl-th)].
%
%10
\bibitem{thoma04} M.H.~Thoma, IEEE Trans. Plasma Science {\bf 32}, 738 (2004)
%
%11
\bibitem{afilinov_jpa03} A. Filinov, M. Bonitz, and W. Ebeling,
% Improved Kelbg potential for correlated Coulomb systems
J. Phys. A {\bf 36}, 5957 (2003).
%

%
%14
\bibitem{kelbg} G.~Kelbg, Ann. Phys. (Leipzig) {\bf 12}, 219 (1962); {\bf 13}, 354 (1963).
%
\bibitem{dusling09} %The idea to use a Kelbg-type effective potential also for %quark matter
%%was independently proposed  by
K.~Dusling and C.~Young, arXiv:0707.2068  [nucl-th].
%However, their potentials are limited to weakly nonideal systems.
%
%%
%%9
\bibitem{Filinov:2009pimc}
% Equation of state of strongly coupled quark-gluon plasma: Path integral Monte Carlo results.
V.S. Filinov, % %
M. Bonitz, Yu. B. Ivanov, {\em et al.},          %V. V. Skokov, P. R. Levashov, and V. E. Fortov,
Contrib. Plasma Phys.,  \textbf{49}, 536 (2009).
%
\bibitem{Filinov:2010pimc}
%Quantum simulations of strongly coupled quark-gluon plasma.
V.S. Filinov, %{\em et al.}, %
M. Bonitz, Yu. B. Ivanov, {\em et al.},        %V. V. Skokov, P. R. Levashov, and V. E. Fortov,
%Talk given at International Workshop on High Density Nuclear Matter, Cape Town, South Africa, 6-9 Apr 2010.
e-Print: arXiv:1006.3390 [nucl-th].
%
% lattice QCD results for EOS
%%\bibitem{peshier02} A.~Peshier, B. K\"ampfer, and G.~Soff, Phys. Rev. D {\bf 66}, 094003 (2002)
%
\bibitem{ColWig11}
%Quantum Color Dynamic Simulations of the Strongly Coupled Quark-Gluon Plasma.
V. S. Filinov, M. Bonitz, Y.B. Ivanov, V.V. {\em et al.} %,Skokov, P.R. Levashov, and V.E. Fortov,
Contrib. Plasma. Phys. {\bf 51}, N4, 322-327 (2011).
%
%21
\bibitem{Lattice02}
% Temporal quark and gluon propagators: Measuring the quasiparticle masses.
P. Petreczky, %{\em et al.}, %
F. Karsch, E. Laermann, {\em et al.},            %S. Stickan, and I. Wetzorke,
% (Bielefeld U.) . Oct 2001. 3pp.
%Contributed to 19th International Symposium on Lattice Field Theory (Lattice 2001), Berlin, Germany, 19-24 Aug 2001.
%Published in
Nucl. Phys. Proc. Suppl. {\bf 106}, 513 (2002).
%e-Print: hep-lat/0110111
%
%22
\bibitem{LiaoShuryak} J. Liao and E. V. Shuryak, Phys. Rev. D \textbf{73}, 014509 (2006).
%
%23
\bibitem{Wong} S. K. Wong, Nuovo Cimento A \textbf{65}, 689 (1970).
%
%24
\bibitem{feynm}  R. P.~Feynman, and A. R.~Hibbs, {\it Quantum
Mechanics and Path Integrals} (McGraw-Hill, New York, 1965).
%
\bibitem{zamalin} V. M.~Zamalin,  G. E.~Norman, and V. S.~Filinov,
{\em The Monte Carlo Method in Statistical Thermodynamics} (Nauka,
Moscow, 1977), (in Russian).
%
%25
\bibitem{filinov_ppcf01} V. S. Filinov, M. Bonitz, W. Ebeling, and V. E. Fortov,
%%Thermodynamics of hot dense H-plasmas: Path integral Monte Carlo simulations and analytical approximations
 Plasma Phys. Control. Fusion {\bf 43}, 743 (2001).
%%
%25
\bibitem{Prosperi}
%On the running coupling constant in QCD
G. M. Prosperi, M. Raciti, and C. Simolo,
Prog. Part. Nucl. Phys. {\bf 58}, 387 (2007). % [arXiv:hep-ph/0607209].

%24
\bibitem{rinton} A. V.~Filinov, V. S.~Filinov, Yu. E.~Lozovik and M.~Bonitz,
{\em Introduction to Computational Methods for Many-Body Physics},
Ed. by M. Bonitz and D. Semkat (Rinton Press, Princeton, 2006).

\end{thebibliography}

\end{document}